\documentclass{elsart4-1}

\usepackage{amssymb}

\usepackage[english,francais]{babel}

\usepackage{amsmath}
\usepackage{graphicx}
\usepackage{color}
\usepackage{bbold}
\usepackage{subcaption} 
\usepackage[noadjust]{cite}

\newtheorem{e-proposition}[theorem]{Proposition}

\newtheorem{e-definition}[theorem]{Definition\rm}

\renewcommand{\theequation}{\arabic{equation}}
\def\og{\leavevmode\raise.3ex\hbox{$\scriptscriptstyle\langle\!\langle$~}}
\def\fg{\leavevmode\raise.3ex\hbox{~$\!\scriptscriptstyle\,\rangle\!\rangle$}}

\newcommand{\be}{\begin{equation}}
\newcommand{\ee}{\end{equation}}

\newcommand{\bra}[1]{\left\langle \, #1 \right|}
\newcommand{\ket}[1]{\left| #1  \right\rangle}

\newcommand{\expec}[1]{\left\langle #1 \right\rangle} 
\newcommand{\comm}[2]{\left[ #1, #2 \right]}
\newcommand{\lind}[1]{\mathcal{D}\left[#1\right]}
\newcommand{\coup}[2]{\mathcal{C}\left[#1, #2 \right]}
\newcommand{\meas}[1]{\mathcal{M}\left[#1\right]}
\newcommand{\Lio}{\mathcal{L}}
\newcommand{\tr}[1]{\text{Tr}\left( #1 \right)}

\begin{document}

\centerline{Physics or Astrophysics/Header}
\begin{frontmatter}


\selectlanguage{english}
\title{Detecting itinerant single microwave photons}


\selectlanguage{english}
\author[chalmers]{Sankar Raman Sathyamoorthy},
\ead{sankarr@chalmers.se}
\author[UQ]{Thomas M. Stace} and
\author[chalmers]{G\"oran Johansson}
\ead{goran.l.johansson@chalmers.se}

\address[chalmers]{Department of Microtechnology and Nanoscience, MC2, Chalmers University of Technology, S-41296 Gothenburg, Sweden}
\address[UQ]{Centre for Engineered Quantum Systems, School of Physical Sciences, University of Queensland, Saint Lucia, Queensland 4072, Australia}


\medskip
\begin{center}
{\small Received *****; accepted after revision +++++}
\end{center}

\begin{abstract}
Single photon detectors are fundamental tools of investigation in quantum optics and play a central role in measurement theory and quantum informatics. Photodetectors based on different technologies exist at optical frequencies and much effort is currently being spent on pushing their efficiencies to meet the demands coming from the quantum computing and quantum communication proposals. In the microwave regime however, a single photon detector has remained elusive although several theoretical proposals have been put forth. In this article, we review these recent proposals, especially focusing on non-destructive detectors of propagating microwave photons.  These detection schemes using superconducting artificial atoms can reach detection efficiencies of 90\% with existing technologies and are ripe for experimental investigations.

{\it To cite this article: S.R. Sathyamoorthy, T.M. Stace ,G.Johansson, C. R.
Physique XX (2015).}

\vskip 0.5\baselineskip

\selectlanguage{francais}
\noindent{\bf R\'esum\'e}
\vskip 0.5\baselineskip
\noindent
{\bf La d\'etection ... }

{\it Pour citer cet article~: S.R. Sathyamoorthy, T.M. Stace , G.Johansson, C. R.
Physique XX (2015).}

 \vskip 0.5\baselineskip
\noindent{\small{\it Mots-cl\'es~:} Mot-cl\'e1~; Mot-cl\'e2~;
Mot-cl\'e3}}
\end{abstract}
\end{frontmatter}


\selectlanguage{english}
\section{Introduction}
In 1905, his \textit{annus mirabilis}, Einstein not only postulated the existence of light quanta (photons) while explaining the photoelectric effect but also gave a theory (arguably the first) of a photon detector \cite{Einstein1905}. In the decades that followed, significant progress were made in designing photon detectors based on several technologies such as photomultiplier tubes, avalanche photodiodes and cryogenic detectors among others. Such detectors are routinely used in experimental setups in applications ranging from spectroscopy to sensors. Recently, there has been a huge drive coming from the field of quantum information processing to push the efficiency of photon detectors to work at the quantum limit. An ideal photon detector is expected to have 100\% efficiency, a very low dark count and a number resolving nature. Such high efficiency single photon detectors are crucial in the implementation of quantum cryptography proposals such as quantum key distribution (QKD) \cite{LoNatPhot2014}, in experimental probing of the foundations of quantum mechanics such as Bell tests \cite{GiustinaNature2013} and in implementing all optical quantum computers \cite{KnillNature2001} among several other applications. Several single photon detectors have been realized in the optical regime \cite{BullerMST2010,EisamanRSI2011} and are part of the standard quantum optics toolkit. However, single photon detectors at microwave frequencies have been difficult to implement and researchers have resorted to special schemes for homodyne and correlation measurements \cite{MariantoniArxiv2005, SilvaPRA2010}.   The focus of this article is to review some of the proposals that have been put forth to fill this gap.

Of particular interest among photodetectors are the non-destructive ones. These detectors are transparent to the incoming photons and the measurement scheme is termed as quantum nondemolition (QND). QND measurements were first proposed to detect gravitational waves by evading the measurement back-action on the system \cite{BraginskySPU1975, BraginskyJETP1977, ThornePRL1978, UnruhPRB1979, BraginskyScience1980}. These measurement schemes were well suited for the field of quantum optics leading to successful implementation of QND measurements of photon flux in the optical regime \cite{GrangierNature1998}. QND detectors play a major role in schemes such as quantum error correction \cite{SteanePRL96}, state-preparation by measurement \cite{RuskovPRB2003, BishopNJP2009} and one way quantum computing \cite{RaussendorfPRL2001}. We will focus on recent proposals for nondestructive detection of microwave photons in section \ref{sec:qnd}.


We have to nevertheless note that, photon detection schemes (including QND detection) have been shown earlier for microwaves stored in cavities \cite{SchusterNature2007, GuerlinNature2007, WangPRL2008, JohnsonNature2010}. While strong interactions could be mediated between the photons in the cavity and matter used as a detector, the use of cavities also complicate the setup as one has to worry about the bandwidth of operation and the compromise between quality factor and reflection. Schemes to catch microwave photons in a cavity by tuning their coupling to transmission line have been developed recently \cite{YinPRL2013, WennerPRL2014, FlurinPRL2015}. However, one has to know the exact shape, width and the arrival time of the wave-packet to completely absorb the photon in to the cavity. These additional problems make such schemes less attractive for applications such as those discussed above. Hence for the purpose of this review, we will focus on detectors designed for propagating microwave photons.  

The current article is organized as follows. In section \ref{sec:cbjj}, we will discuss   proposals for microwave photon detection that are destructive in nature. These proposals are based on current biased Josephson junctions (CBJJ). In the next section, we will look at proposals that use the photon-photon interaction mediated using a superconducting artificial atom to perform QND detection of microwave photons. In the final section, we will summarize the main messages from these proposals.

\section{Photon detectors based on Josephson junctions}
\label{sec:cbjj}
\begin{figure}[]
\centering
\includegraphics[width=0.7\textwidth]{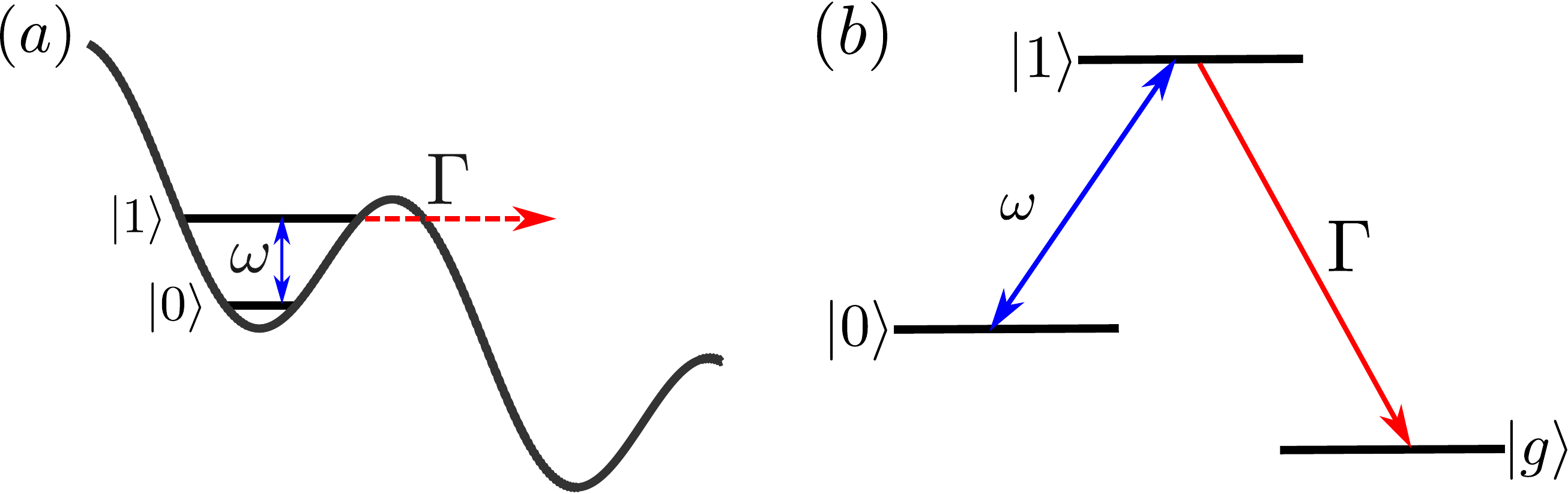}
\caption{(a) Tilted washboard potential of a current biased Josephson junction. The bias current $I_b$ is chosen such that only two energy levels exist within each well. (b) A three level $\Lambda$ system that maps to the spectrum shown in (a).}
\label{fig_washboard_lambda}
\end{figure}

Initial proposals for detecting propagating microwave photons were based on current biased Josephson junctions (CBJJ) \cite{RomeroPRL2009, RomeroPhysScr2009, PeropadrePRA2011}. These junctions have a potential energy $U(\delta) = -(I_c \Phi_0/2\pi) \cos(\delta) - I_b \delta$, where $I_c$ is the critical current of the junction, $\Phi_0 = h/2e$ is the flux quantum and $\delta$ is the superconducting phase difference across the junction. Such a potential, known as a tilted washboard potential (see Fig.\ref{fig_washboard_lambda}(a)), has several local minima or potential wells that have discrete set of states. We can tune the potential barrier of the local minima and the number allowed energy levels using the bias current $I_b$. For a particular value of $I_b$ (depending on $I_c$), each of the wells acts as a two level system that can be used as a qubit. The lifetime of these levels $\ket{0}$ and $\ket{1}$ depend on the tunneling distance to the continuum of modes to the right. As seen from the figure, the tunneling from $\ket{0}$ is exponentially suppressed compared to the decay from the state $\ket{1}$. If we neglect the tunneling from the state $\ket{0}$, the setup can be mapped to a three level $\Lambda$ system (Fig.\ref{fig_washboard_lambda}(b)), with $\ket{1}$ being a metastable state with a short lifetime set by the rate $\Gamma$. The incoming photon with frequency close to the transition frequency $\omega$ excites the qubit to state $\ket{1}$. The excitation from $\ket{1}$ decays irreversibly with the rate $\Gamma$ into the continuum that is labeled here as the state $\ket{g}$. This leads to a voltage drop across the junction that can be measured classically and the excitation is lost making this a destructive scheme.

The performance of the detector is characterized by the probability of absorption of the incoming photon. This is theoretically calculated as follows. We start with the non-Hermitian Hamiltonian of the $\Lambda$ system in the real space representation ($\hbar=1$)
\be
H=\left(\omega - i \Gamma/2 \right) \ket{1}\bra{1} + i v_g \int dx \left(\psi_L^\dagger \partial_x \psi_L - \psi_R^\dagger \partial_x \psi_R  \right) + V \int dx \; \delta(x) \left[ (\psi_R+\psi_L)\ket{1}\bra{0} + h.c. \right],
\ee
where $\psi_{R/L}$ is the radiation field traveling right/left with a group velocity $v_g$. The interaction between the photons and the $\Lambda$ system is modeled as a delta potential of strength $V$ at $x=0$. A general single excitation wave-function of the total system is of the  form
\be
\ket{\phi} = \int dx \left[ \xi_R(x,t) \psi_R^\dagger(x) + \xi_L(x,t) \psi_L^\dagger(x) \right] \ket{0,\text{vac}}  + e(t) \ket{1,\text{vac}}   
\ee
where $\ket{0,\text{vac}}$ is the state with the $\Lambda$ system in state 0 and the radiation field in vacuum. The single photon wave-packets are given by $\xi_{R/L}(x,t)$. Solving the Schr\"odinger equation for the above Hamiltonian with this state, we get three coupled equations for the coefficients $\xi_R(x,t)$, $\xi_L(x,t)$ and $e(t)$.  From the solution of these coupled equations the probability of level $\ket{g}$ can be calculated as $P_g = 1 - ||\phi||^2$. The value of this probability at long times is taken as the measure the detector efficiency. With the $\Lambda$ system in the middle of an open transmission line, the maximum attainable efficiency was calculated to be 50\% \cite{RomeroPRL2009}. This could be improved by having several such scatterers along the line which also increased the bandwidth of operation. In \cite{PeropadrePRA2011}, it was theoretically shown that by placing one atom at the end of transmission line (in front of a mirror), the efficiency could be increased to 100\%.

The above discussed analysis were done for an ideal $\Lambda$ system, of which the CBJJ is only an approximation. While the tunneling from the state $\ket{0}$ is much smaller than the tunneling from the state $\ket{1}$, it is not zero. The direct tunneling from the ground state $\ket{0}$ leads to dark counts, which affects the performance of the setup. The performance of CBJJ as a photodetector (referred as Josephson photomultiplier (JPM) in the following references to distinguish from phase qubits which ideally work at different parameter regime) was analyzed both experimentally and theoretically including the direct tunneling from the state $\ket{0}$ in references \cite{ChenPRL2011, PoudelPRB2012, GoviaPRA2012, GoviaPRA2014}. However, these works are primarily aimed at measuring the photon occupation number inside a cavity. Further analysis is needed to check the efficiency of such a system to be used as a detector for itinerant single photons. 

\section{QND detection of itinerant microwave photons}
\label{sec:qnd}
\subsection{A transmon in a transmission line}
The transmon \cite{KochPRA2007}, a widely used superconducting artificial atom, is a single cooper pair box shunted with a capacitance $C_S$. The large shunt capacitance reduces the charging energy $E_C$ of the circuit that also reduces the sensitivity of the qubit to charge noise and is one of the main reasons for the transmon's popularity. Recently, it was shown that a transmon could behave as a Kerr medium that imparts a giant conditional phase shift on microwave fields incident on the atom \cite{IoChunPRL2013}. Following schemes for QND detection of photons based on the cross-Kerr effect in the optical regime, proposals for nondestructive single photon detection in the microwave frequencies using the transmon were analyzed in references \cite{BixuanPRL2013,SankarPRL2014,BixuanPRB2014}. We will briefly review these setups in this section.

\begin{figure}
\centering
\includegraphics[scale=0.65]{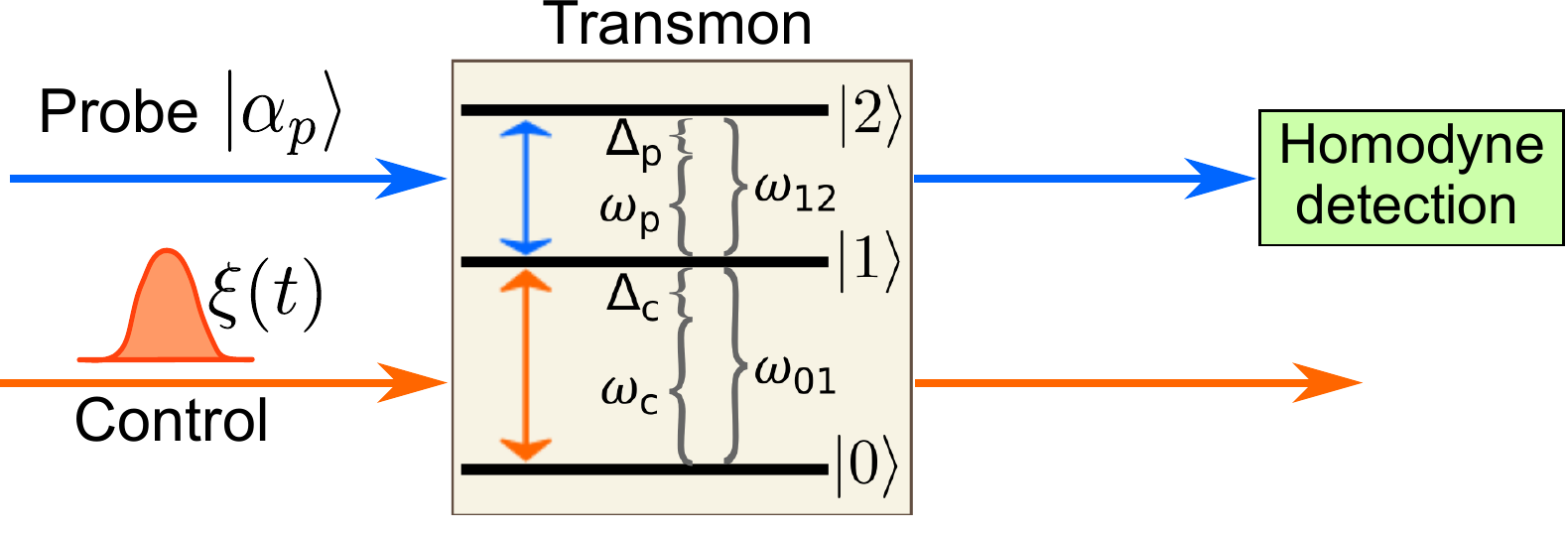}
\caption{A scheme for single photon detection using a single three level system. Such a setup could be achieved in superconducting circuits with a transmon at the end of a transmission line. The difference in the homodyne current with and without a single control photon in the wave-packet $\xi(t)$ constitutes the signal.}
\label{fig_single_transmon}
\end{figure}

For the purpose of this article, we will consider the transmon as essentially a three level system with the Hamiltonian ($\hbar=1$)
\be
H = -\omega_{01} \ket{0}\bra{0} + \omega_{12} \ket{2}\bra{2},
\ee
where $\omega_{ij} = E_j - E_i$ is the energy difference between levels $i$ and $j$. The $1-2$ transition is driven by a coherent probe field with frequency $\omega_p$ and amplitude $\alpha_p$. The $0-1$ transition is driven by the control field of frequency $\omega_c$ which contains either 0 or 1 photon (See Fig. \ref{fig_single_transmon}). In the rotating frame of the input fields, the Hamiltonian can be written including the coherent drive as
\be
H = -\Delta_c \ket{0}\bra{0} + \Delta_p \ket{2}\bra{2} + \Omega_p(L_{12}+L_{21}),
\ee
where the detunings $\Delta_c=\omega_{01}-\omega_c$, $\Delta_p=\omega_{12}-\omega_p$ and we have taken $\alpha_p = i \Omega_p$, with a real $\Omega_p$. We will also denote the coupling operators as $L_{ij} = \sqrt{\Gamma_{ij}} \ket{i}\bra{j}$ where $\Gamma_{ij}$ is the decay rate from $j$th energy level to $i$th level. The above Hamiltonian is valid for an atom with one input-output port. Such a setup can be achieved by placing the transmon at the end of a semi-infinite transmission line. With the transmon initially in the ground state $\ket{0}$, the probe field does not interact with the $1-2$ transition. The probe field is scattered only when the control field has a photon which excites the transmon to the state $\ket{1}$. By monitoring the output probe field using homodyne detection the presence of the single control photon can thus be inferred. 

The single photon in the control field can be modeled as the output from a fictitious cavity with damping rate $\kappa(t)$. By modulating $\kappa(t) = \xi(t)/\sqrt{\int_t^\infty |\xi(s)|^2ds}$, the shape of the temporal wave packet $\xi(t)$ containing the single photon can be arbitrarily set \cite{Gough2012}. The master equation of the setup is
\begin{align}
\label{Eq_MECavity}
\dot{\rho} &= -i\comm{H}{\rho} + \left(\lind{L_{01}} + \lind{L_{12}} \right)\rho + \kappa(t) \lind{a} \rho - \sqrt{\kappa(t)}\; \coup{a}{L_{01}}\rho \nonumber \\
&\equiv \Lio_{\text{cav}}\rho, 
\end{align}
where $\rho$ is the total density matrix of the source cavity and the transmon, $a$ ($a^\dag$) is the annihilation (creation) operator for the photons in the cavity and the dissipation super-operator is given by $\lind{c} \rho = c\rho c^{\dagger }-\frac{1}{2}c^{\dagger }c\rho -\frac{1}{2} \rho c^{\dagger }c$. We have also defined a Liouvillian $\Lio_{cav}$ and a coupling super-operator $\coup{c_1}{c_2} \rho = \left[c_2^{\dagger }, c_1\rho \right]+\left[\rho~c_1^{\dagger },c_2\right]$ for shorthand.


The output probe field that is measured using the homodyne detector is given by the standard input-output formalism as $\Omega_{p, out} = \Omega_p + L_{12}$. However, we are only interested in the change of the probe field with and without a single photon in the control field, i.e. only on  $L_{12}$. The interesting part of the homodyne current can thus be defined as
\be
j(t) dt = \sqrt{\eta}\expec{e^{i\phi}L_{12}(t)+e^{-i\phi}L_{21}(t)} dt + dW(t), 
\ee
where $0 \leq \eta \leq 1$ is the efficiency of the homodyne detector and $\phi$ is the phase of the local oscillator that specifies the quadrature of the measurement. $dW(t)$ is a Wiener increment with a mean $E[dW(t)]=0$ and variance $E[dW(t)^2]=dt$ where $E[\cdot]$ is the ensemble average. To convert the above time trace of the homodyne current into a binary flag, we define a signal $S=\int_{t_i}^{t_f} j(t) f(t) dt$, where $t_m = t_f-t_i$ is the measurement time window and $f(t)$ is a linear filter function. We will initially take $f(t)$ to be just a square pulse with value 1 when $t_i \leq t \leq t_f$ and 0 otherwise.

To characterize the performance of this setup as a photon detector, we can define the signal to noise ratio (SNR) as
\be
\label{SNR_def}
\text{SNR}=\frac{E[S_1]-E[S_0]}{\sqrt{\text{Var}[S_1]+\text{Var}[S_0]}}, 
\ee
where $S_{0/1}$ is the signal with 0 or 1 photon in the control field and $\text{Var}[X]=E[X^2]-E[X]^2$ is the variance. The SNR can be calculated from the above master equation \eqref{Eq_MECavity}. The average and variance in  this case are
\begin{align}
E[S_0] &=0 \nonumber \\
E[S_0^2]&=t_m \nonumber \\
E[S_1]&=\sqrt{\eta}\int_{t_{i}}^{t_{f}} \expec{\hat{y}} dt \nonumber \\
E[S_1^2] &=  \int_{t_{i}}^{t_{f}} dt_1 \int_{t_{i}}^{t_{f}} dt_2 \;E[j(t_1) j(t_2)].
\end{align}
The two time correlation function can be evaluated using the quantum regression theorem \cite{Lax66} as
\begin{align}
E[j(t_1) j(t_2)] &= \Theta(t_2-t_1)\bigg(\eta \tr{(e^{i\phi}L_{12} + e^{-i\phi} L_{21})\mathcal{T}(t_2-t_1)(e^{i\phi} L_{12} \rho(t_1) + e^{-i\phi} \rho(t_1) L_{21})} +\delta(t_2-t_1)\bigg) \nonumber\\ &+ \Theta(t_1-t_2)\bigg(\eta \tr{(e^{i\phi}L_{12} +e^{-i\phi} L_{21})\mathcal{T}(t_1-t_2)(e^{i\phi}L_{12} \rho(t_2) +e^{-i\phi} \rho(t_2) L_{21})} +\delta(t_1-t_2)\bigg)
\end{align}
where $\mathcal{T}(t_2-t_1) \mathbb{Y}(t_1) = \mathbb{Y}(t_2)$ with $\mathbb{Y}(t)=e^{i\phi}L_{12} \rho(t) + e^{-i\phi} \rho(t) L_{21}$. The time evolution operator $\mathcal{T}(t_2-t_1)$ is evaluated by solving $\dot{\mathbb{Y}} = \Lio_{\text{cav}}\mathbb{Y}$. By definition, the step function $\Theta(t)=0$ for $t<0$ and 1 otherwise. 

The signal to noise ratio is an ideal measure for Gaussian statistics but is not necessarily a good measure for other distributions that need higher orders moments for complete description. In order to collect such statistics of the signal distribution and define more relevant measures, we use the formalism of stochastic master equations(SME). These equations describe the evolution of the system under measurements and can be considered as an unravelling of the average system dynamics described by the above master equations \cite{CarmichaelQuantumOptics}. With the tunable cavity as the photon source, the SME is given by
\be
\label{Eq_SMECavity}
d\rho = \Lio_{\text{cav}}\rho dt  + \sqrt{\eta} \meas{\Lambda_{12}}{\rho} \; dW(t),
\ee
where we have defined a measurement super-operator $\meas{c}{\rho} = (e^{i\phi}c\rho + e^{-i\phi}\rho c^{\dagger })-\expec{e^{i\phi}c+ e^{-i\phi} c ^{\dagger }}{\rho}$, that describes the back-action of the measurement on the evolution of the system. 


To get the distribution of the signal, we numerically solve the above equations with $n=0$ and $n=1$ photon in the control field. Each run of this simulation is called a trajectory which gives a particular value of $S$. The number of trajectories should be high enough to get a probability distribution for $S$. In this case, we can define fidelity of single photon detection as
\be
F = P(0|S \leq S_0^T)P(S \leq S_0^T) + P(1 | S \geq S_1^T)P(S \geq S_1^T),
\ee 
where $P(0|S \leq S_0^T)$ is the conditional probability to have 0 photons in the control field given that the measured  integrated current $S$ was found to be less than or equal to  a predefined threshold $S_0^T$. $P(1|S \geq S_1^T)$ is defined analogously.

\begin{figure}
\centering
\includegraphics[width=0.5\textwidth]{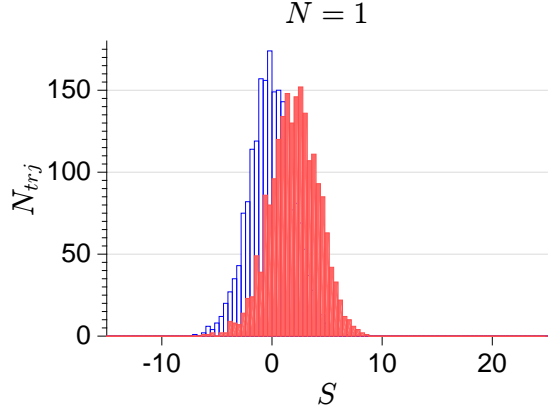}
\caption{Histogram of the integrated homodyne current $S$ with(red) and without(blue) a photon in the control field, using a single transmon ($N=1$) at the end of a transmission line. The number of trajectories in both the cases ($n=0$ and $n=1$ control photons) was 2000. The parameters used in the units of $\Gamma_{01}$ are $\Delta_{01}=\Delta_{12}=0, \Omega_p=0.35$ and $\Gamma_{12}=2$ while the quadrature of measurement is set with $\phi=\pi/2$. The input control photon is of Gaussian temporal shape with $\Gamma_{ph}=0.8$ and $T_{ph}=4$. The time of measurement $t_m$ was optimized to give the best SNR. As is evident from the distribution, the SNR in this case is less than 1.}
\label{fig_hist_1tr}
\end{figure}

As first shown in \cite{BixuanPRL2013}, the signal generated using a single transmon in the above setup cannot overcome the quantum noise. This can be seen from the distribution of the signal shown in Fig.\ref{fig_hist_1tr}, which is obtained from the numerical simulation of the above stochastic master equations. The distribution is obtained from 2000 trajectories for $n=0$ and $n=1$ control photons, assuming no additional losses such as dephasing in the transmon and with perfect homodyne detection ($\eta=1$). In this simulation, we have also taken the single photon to be in a Gaussian wave-packet given by 
\[\xi(t) = \left(\frac{\Gamma_{ph}^2}{2\pi} \right)^{1/4} \exp\left( \frac{-\Gamma_{ph}^2(t-T_{ph})^2}{4}\right),\]
where $\Gamma_{ph}$ gives the width of the wave packet and $T_{ph}$ is the time of arrival of the peak of the wave packet to the transmon, with the normalization condition $ \int |\xi(t)|^2 dt = 1$. The parameter range is chosen to give optimal SNR which was calculated to be around 0.7 and correspondingly the fidelity was found to be around 70\%.

\begin{figure}
\centering
\begin{subfigure}[]{0.45\textwidth}
\includegraphics[width=\textwidth]{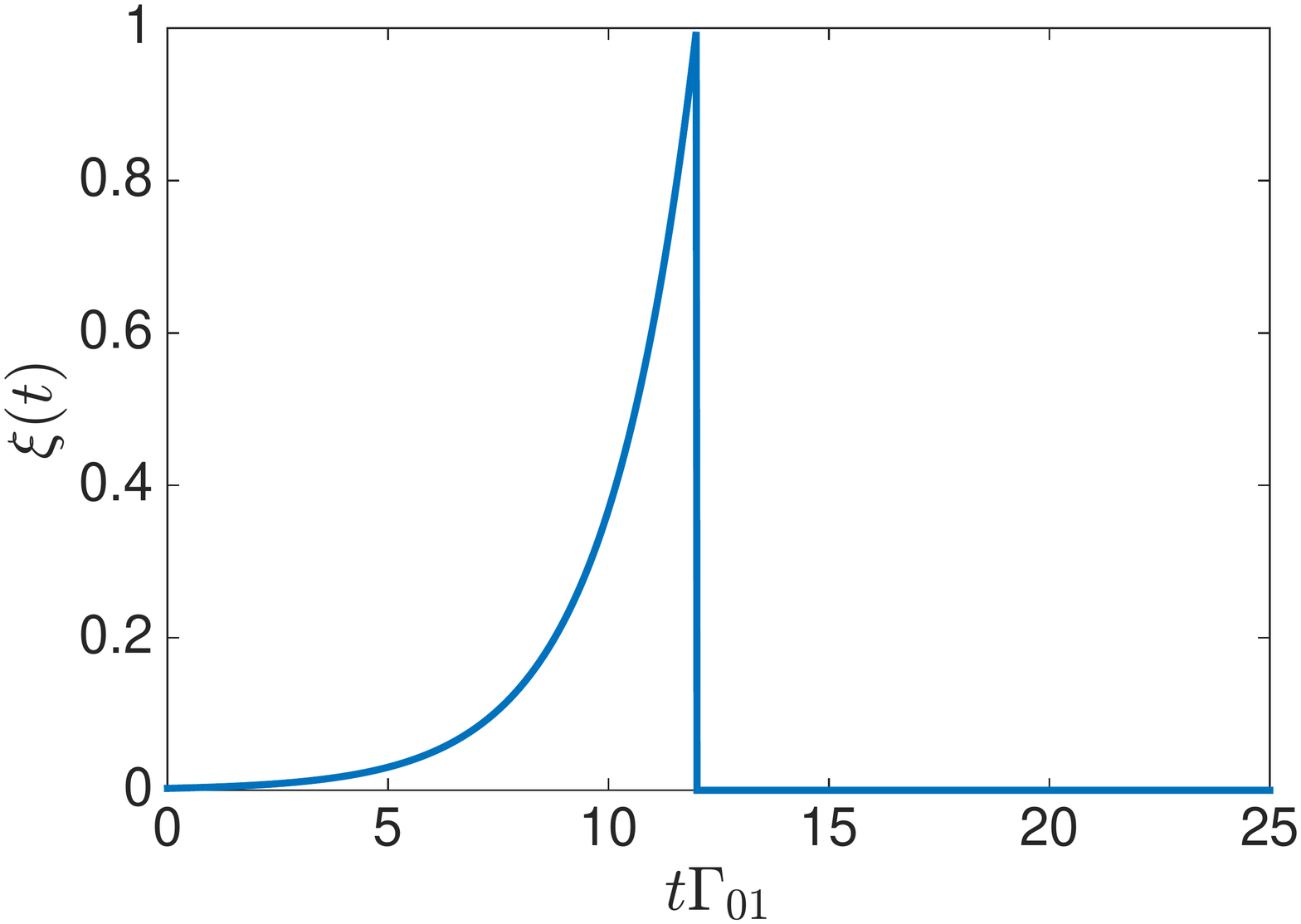}
\end{subfigure}
\begin{subfigure}[]{0.45\textwidth}
\includegraphics[width=\textwidth]{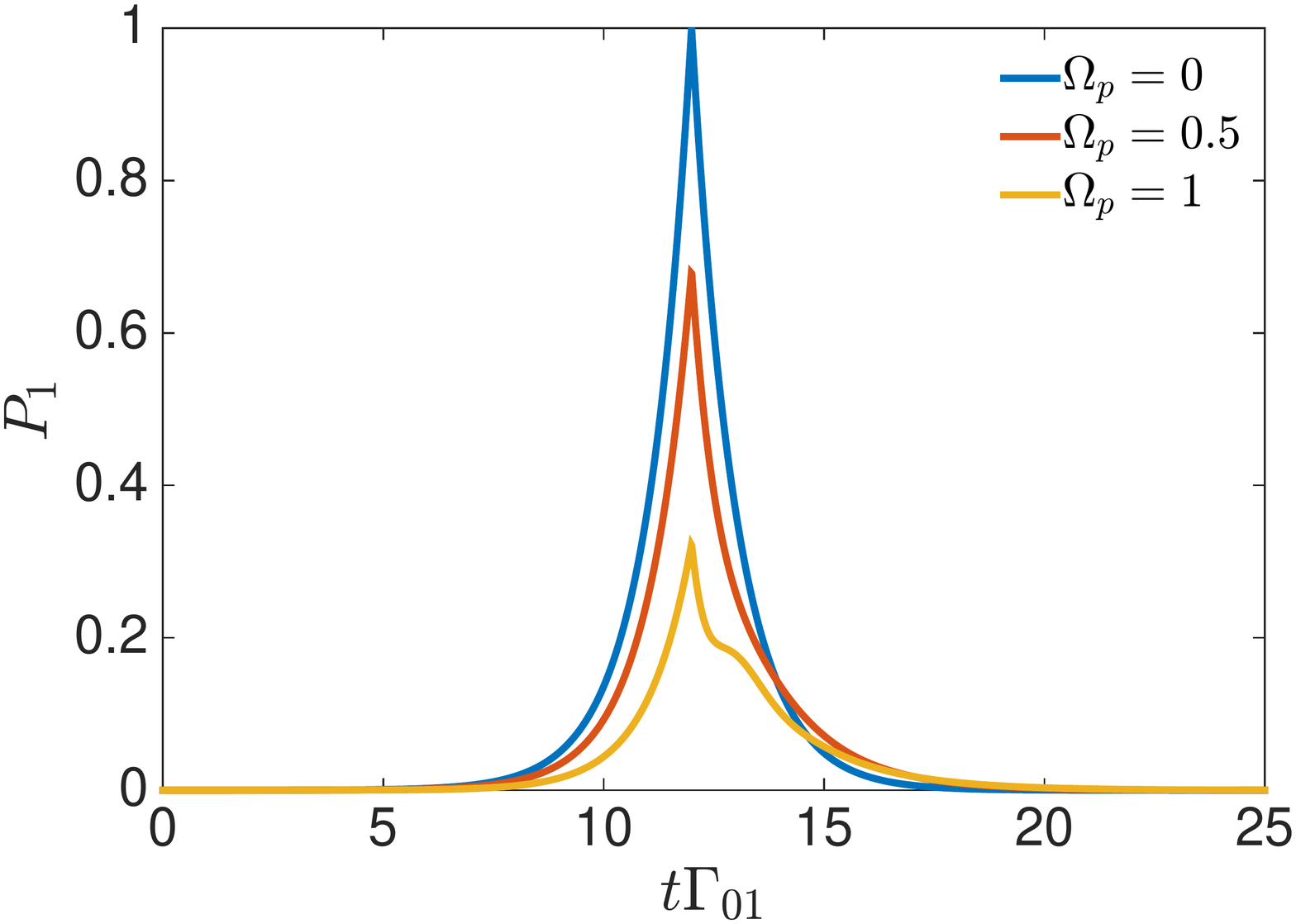}
\end{subfigure}
\caption{Excitation probability of the $0-1$ transition of a transmon: In the absence of a coherent probe field on resonance with $1-2$ transition, the single photon in the control field is completely absorbed by the transmon, if we enclose it in a rising exponential wave-packet $\xi(t) = \Theta(t-T_{ph})\sqrt{\Gamma_{ph}} \exp\left(\frac{\Gamma_{ph} t}{2}\right)$ with $\Gamma_{ph}=\Gamma_{01}$. Switching on the probe field, reduces the maximum excitation probability as can be seen in the second panel.}
\label{fig_Pexc_vs_Op}
\end{figure}

The reason for $SNR < 1$, can be understood heuristically as follows. A single atom can process only one excitation of a transition per interaction time. Even if we consider that the control photon is completely absorbed by the atom, it can only scatter a single photon from the coherent probe field. As the noise of the coherent field is of the order of a photon, this displacement in its amplitude is not enough to distinguish between the cases of having $n=0$ and $n=1$ control photons. In fact, it can be seen that driving the $1-2$ transition with the probe field actually reduces the excitation probability of the $0-1$ transition (i.e. the incoming single photon is not completely absorbed by the atom) and thus we don't even reach the limit discussed above (See Fig.\ref{fig_Pexc_vs_Op}).

\subsection{Beyond a single transmon}
In order to overcome the limitations as discussed above, we look at using more than 1 transmon to displace the coherent probe over the limit set by the quantum noise. As a first step in this regard, we briefly review a convenient tool known as the $(S,L,H)$ formalism \cite{GoughCommMathPhys2009,GoughIEEE2009} to derive the master equations for connected quantum systems. In this formalism, each subsystem is described by a triplet $G \equiv (S,L,H)$, where $S$ is the scattering matrix, $L$ is the vector of coupling operators and $H$ is the Hamiltonian of the subsystem. Once the triplets are identified for each of the subsystem, the total triplet for the composite system can be written using the following products.

The series product $\triangleleft$ of the triplets describes feeding the output from one subsystem into another
\be
G_2 \triangleleft G_1 = \Bigg(S_2 S_1, S_2L_1 + L_2,  H_1 + H_2 + \frac{1}{2i}\left(L_2^\dag S_2 L_1 - L_1^\dag S_2^\dag L_2 \right) \Bigg).
\label{seriesproduct} 
\ee

The concatenation product $\boxplus$ is used for composing subsystems into a system with	 stacked channels
\be
G_2 \boxplus G_1 = \left( \begin{pmatrix} S_2 & 0 \\ 0 & S_1 \end{pmatrix}, \begin{pmatrix} L_2 \\  L_1 \end{pmatrix}, H_2 + H_1 \right).  
\label{catproduct} 
\ee

The $(S,L,H)$ formalism can also handle feedbacks. However, as we won't use them in this particular review we refer the readers interested in feedbacks to the above references. 

Using the above defined products, we can write down the $(S,L,H)$ triplet for the whole system 
\be
G_{\rm tot} = \left(S_{\rm tot}, \begin{pmatrix} L_1 \\ \vdots \\ L_n \end{pmatrix}, H_{\rm tot}\right),
\ee
from which we can extract the corresponding master equation as 
\be
\label{Eq:MEfromSLH}
\dot{\rho} = -i\comm{H_{\rm tot}}{\rho} + \sum_{i=1}^n \lind{L_i}\rho.
\ee
The output from the $i^{th}$ channel is just given by $L_i$. The $(S,L,H)$ triplet for the case of a single transmon with three levels in front of a mirror is
\begin{align}
G_\text{tr} = \left(
{\mathbb{1}}_2,
\begin{pmatrix}
L_{01} \\
L_{12} 
\end{pmatrix} ,
H_\text{tr}
\right),
\end{align}
with the same Hamiltonian as in the previous section. The $(S,L,H)$ triplets for the tunable cavity and the coherent probe in their corresponding rotating frames are
\be
G_{\text{cav}} = (1,\sqrt{\kappa(t)}a,0)
\ee
and
\be
G_{\alpha_p} = (1,\alpha_p,0). 
\ee

\begin{figure}
\centering
 \includegraphics[scale=0.5]{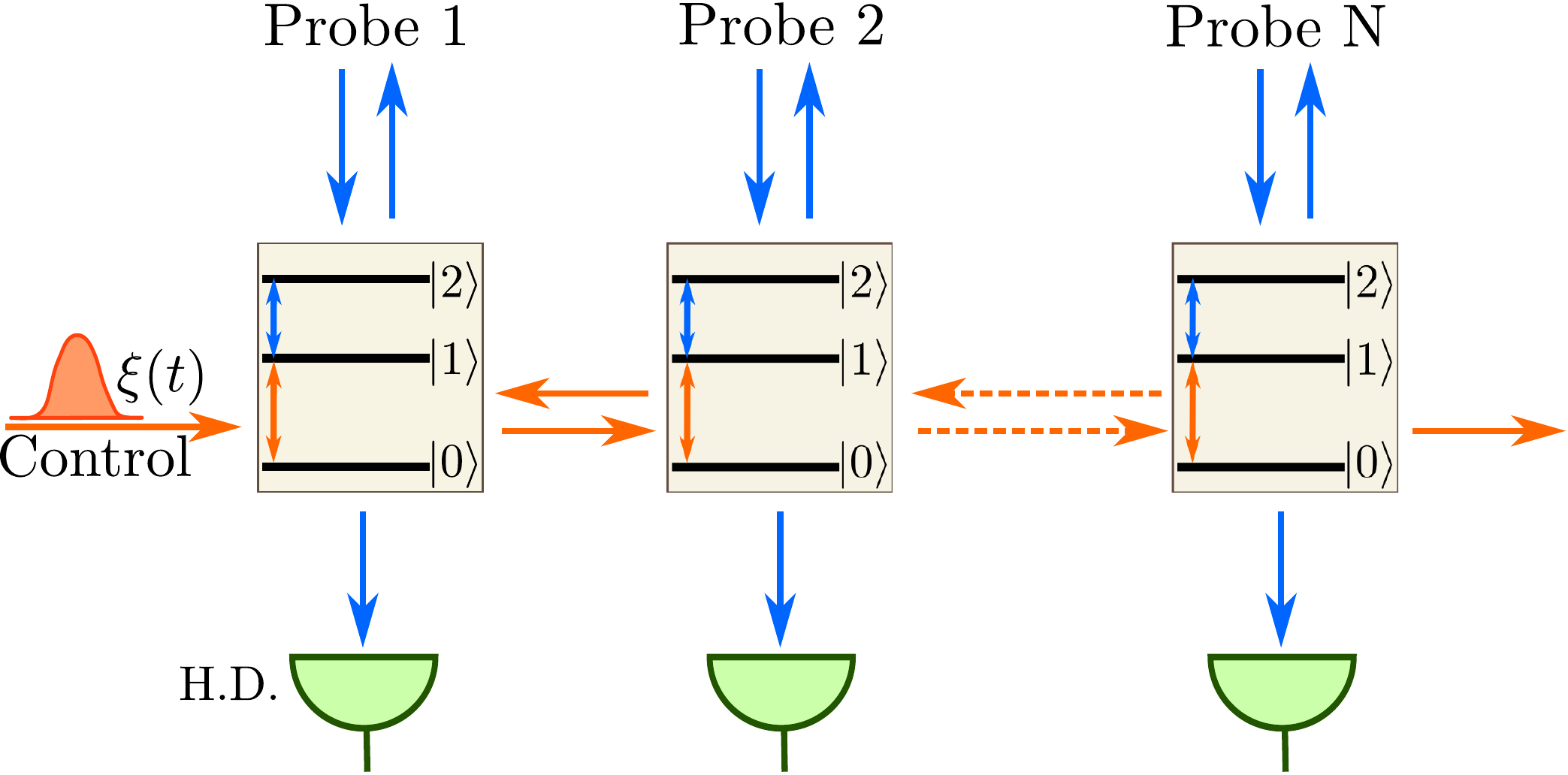}
 \caption{Schematic setup of $N$ transmons coupled to a transmission line. Each of the transmons is probed by a coherent field and the change in the reflected or transmitted amplitude is measured by a homodyne detector.} 
 \label{fig_coupled_transmons}
\end{figure}

With these triplets, we are now ready to look at different approaches with which we can connect different transmons. A first approach that is experimentally straightforward to implement is to couple many transmons to a transmission line one after the other (see Fig. \ref{fig_coupled_transmons}). One can imagine two different scenarios in this case. The distance between the transmons could be much less than the wavelength of the incoming fields or it could be several factors higher. Both of theses scenarios were analyzed in \cite{BixuanPRL2013} and found to be not helpful for the problem of single photon detection. The heuristic understanding for these is as follows. In case the transmons are closer than the transition wavelengths, the energy levels hybridize to form a larger atom with a different normalized coupling. In this case we are back to the case of a single transmon, albeit with a different set of parameters. While this limitation could in principle be overcome by separating the transmons over larger distances, a Kramers-Kronig type relation between the interaction and losses prevents any useful gain in the signal. To go beyond these limitations, we need a unidirectional coupling between the transmons i.e. a  cascaded setup as discussed in \cite{SankarPRL2014}.

\begin{figure}[b]
\centering
\includegraphics[scale=0.5]{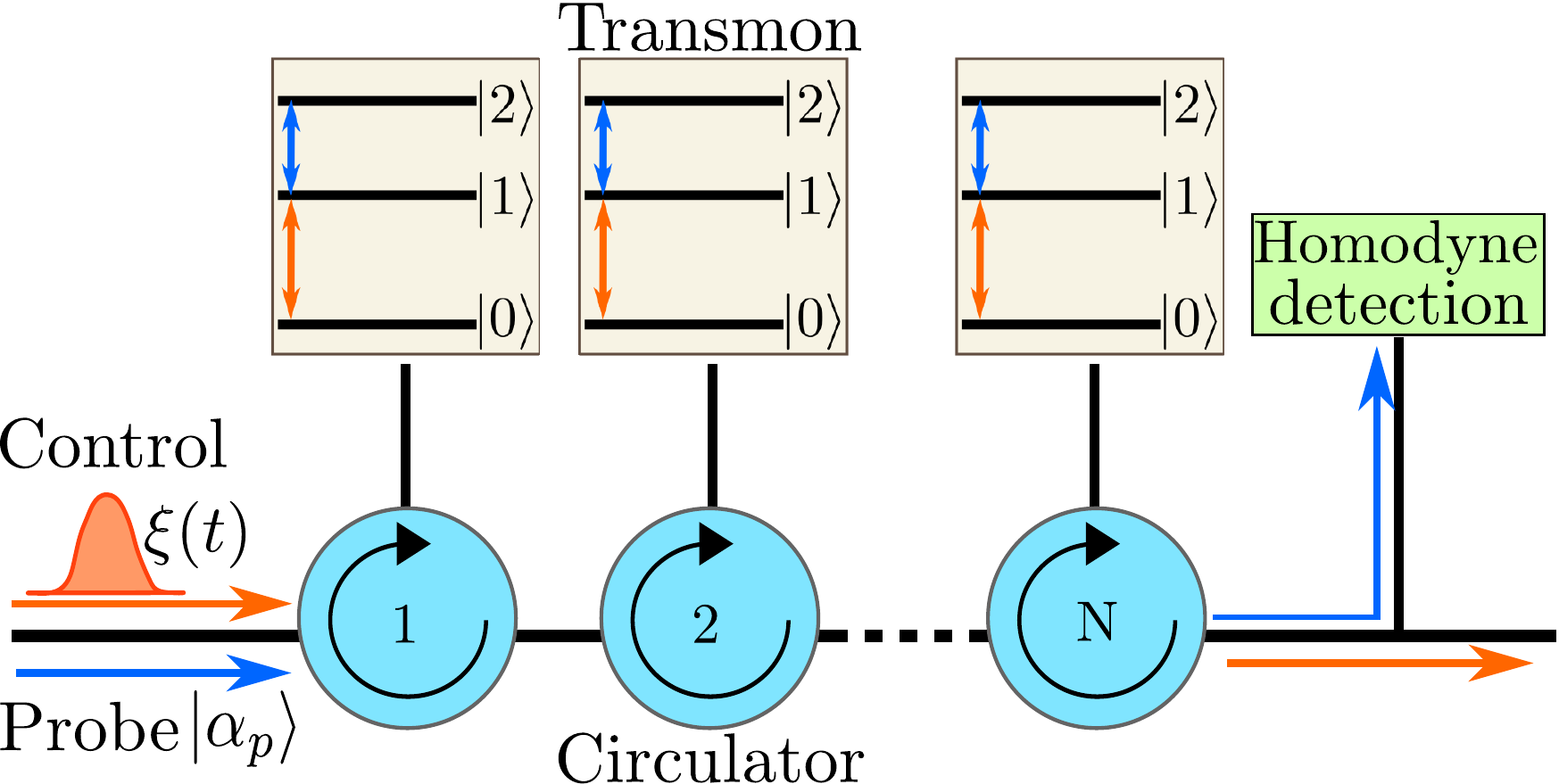}
 \caption{Schematic setup of cascaded transmons for QND detection of microwave photons. Placing the transmons at the end of transmission lines and connecting them via circulators makes this a cascaded system where the fields travel unidirectionally.}
 \label{fig_cascaded_transmons}
\end{figure}

The setup considered is schematically shown in Fig. \ref{fig_cascaded_transmons}. We achieve unidirectional flow of the fields by placing the transmons at the end of transmission lines (similar to having atoms in front of mirrors) and connecting them using microwave circulators. The master equation for this setup is derived as follows. Following the product rules in Eqs. \ref{seriesproduct} and \ref{catproduct}, the $(S,L,H)$ triplet for the setup consisting of $N$ cascaded transmons with a tunable cavity as the photon source and a coherent probe of strength $\alpha_p$ is
\begin{align}
G_\text{tot} &=  G_\text{tr}^{(N)} \triangleleft \ldots \triangleleft G_\text{tr}^{(k)} \triangleleft \ldots \triangleleft G_\text{tr}^{(2)} \triangleleft G_\text{tr}^{(1)} \triangleleft
 \left(G_\text{cav} \boxplus G_{\alpha_p} \right) \nonumber\\
 & = {\left(
{\mathbb{1}}_2,
\begin{pmatrix}
\sqrt{\kappa(t)}a + \Lambda_{01}\\
\alpha_p + \Lambda_{12}
\end{pmatrix} ,
H_\text{tot}
\right),}
\end{align}
where
\begin{align}
H_\text{tot} &= \sum_{j=1}^{N} H_\text{tr}^{(j)} + \frac{1}{2i}\sqrt{\kappa(t)}\left( \Lambda_{10} a - a^\dag \Lambda_{01} \right) + \frac{1}{2i}\left(\alpha_p \Lambda_{21} - \alpha_p^* \Lambda_{12} \right) \nonumber\\
&+ \frac{1}{2i}\sum_{j=1}^N \left(\sum_{k=j+1}^N \left(L_{10}^{(k)} L_{01}^{(j)}  - L_{10}^{(j)} L_{01}^{(k)} \right)\right) + \frac{1}{2i}\sum_{j=1}^N\left(\sum_{k=j+1}^N \left(L_{21}^{(k)} L_{12}^{(j)} - L_{21}^{(j)} L_{12}^{(k)} \right)\right)\nonumber\\
\end{align}
and the collective operators $\Lambda_{ij}=\sum\limits_{k=1}^{N} L_{ij}^{(k)}$. 
This gives the master equation (after some algebra) as
\begin{align}
\dot{\rho} &= -i\comm{H_\text{eff}}{\rho} + \sum_{j=1}^N\left(\lind{L_{01}^{(j)}} + \lind{L_{12}^{(j)}} \right)\rho  + \kappa(t) \lind{a} \rho - \sqrt{\kappa(t)}\; \coup{a}{\Lambda_{01}} \rho \nonumber \\
&- \sum_{j=1}^N \sum_{k=j+1}^N \bigg(\coup{L_{01}^{(j)}}{L_{01}^{(k)}} + \coup{L_{12}^{(j)}}{L_{12}^{(k)}}   \bigg)\rho 
\end{align}
where the effective Hamiltonian is $H_\text{eff} = \sum\limits_{k=1}^{N} H^{(k)}$ with $\text{H}^{(k)}= -\Delta^{(k)}_{01}\ket{0}\bra{0}^{(k)}+\Delta^{(k)}_{12}\ket{2}\bra{2}^{(k)} + \Omega_p(L^{(k)}_{12} + L^{(k)}_{21} )$. We have once again chosen the normalization for the probe field such that $\alpha_p=\Omega_p e^{i\pi/2}$, with $\Omega_p$ being a real number. A master equation using the Fock state formalism could also be derived similarly.

\begin{figure}
\centering
\includegraphics[width=0.7\textwidth]{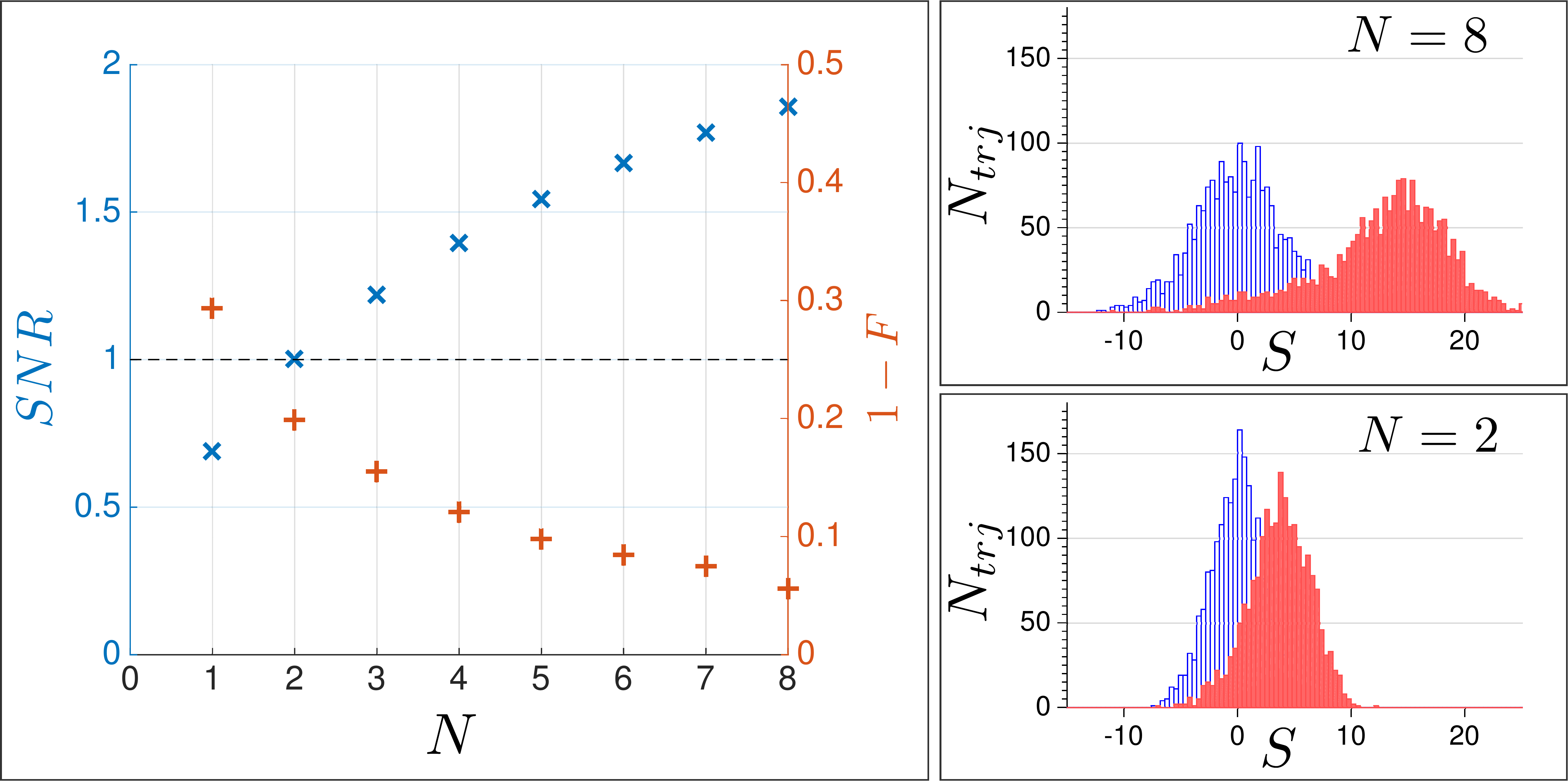}
 \caption{Signal to noise ratio ($SNR$) and the fidelity of photon detection $F$ as a function of $N$, the number of cascaded transmons using the Fock state master equation. Both the control and the probe fields were taken to be on resonance with all of the transmons. The coupling of the individual transmons were tuned to optimize the $SNR$ (Refer \cite{SankarPRL2014} for full parameter list). The input photon was taken to be in a Gaussian wave packets. The insets show the signal distribution for $N=2$ and $N=8$ transmons. Note that for the case with an input control photon $n=1$, the distributions deviate from a normal distribution with the increasing number of transmons, suggesting a memory effect. In such cases, we believe the fidelity of photon detection becomes a better measure than $SNR$.}
 \label{fig_cascaded_results}
\end{figure}

We once again numerically simulate the above equations and corresponding stochastic master equations to calculate the signal-to-noise ratio and the fidelity of photon detection. The main results are shown in Fig. \ref{fig_cascaded_results}. As we can see from the figure, the unidirectional coupling helps accumulate the effects from each of the transmon and we can break-even the noise limit with $N=2$ transmons. The resulting $SNR$ can be fit to a simple $\sqrt{N}$ curve, which would be the expected behavior if we considered each of the scattering events to be independent of one another. The results in Fig. \ref{fig_cascaded_results} are once again for the ideal case of no dephasing in the transmons and with perfect homodyne detection. The setup also seems to work if we take into account some deviation from the ideal setup, though with reduced fidelities. We will not reproduce these results here and would refer the reader interested in full details to reference \cite{SankarPRL2014}.

One of the main concerns in experimentally implementing the above setup is the need of circulators. Currently, these are bulky devices that are lossy and are off-chip. On chip circulators \cite{KochPRA2010,KerckhoffArxiv2015,SliwaArxiv2015} or other unidirectional wave-guides (\'a la the quantum hall edge channels \cite{StacePRL2004}) will make this scheme more attractive to implement. The performance of these setups could be further improved by using a cavity for the probe as shown in \cite{BixuanPRB2014}. We will briefly review this in the next section. 

\subsection{Back to cavities}
\begin{figure}[]
\centering
\includegraphics[width=0.6\textwidth]{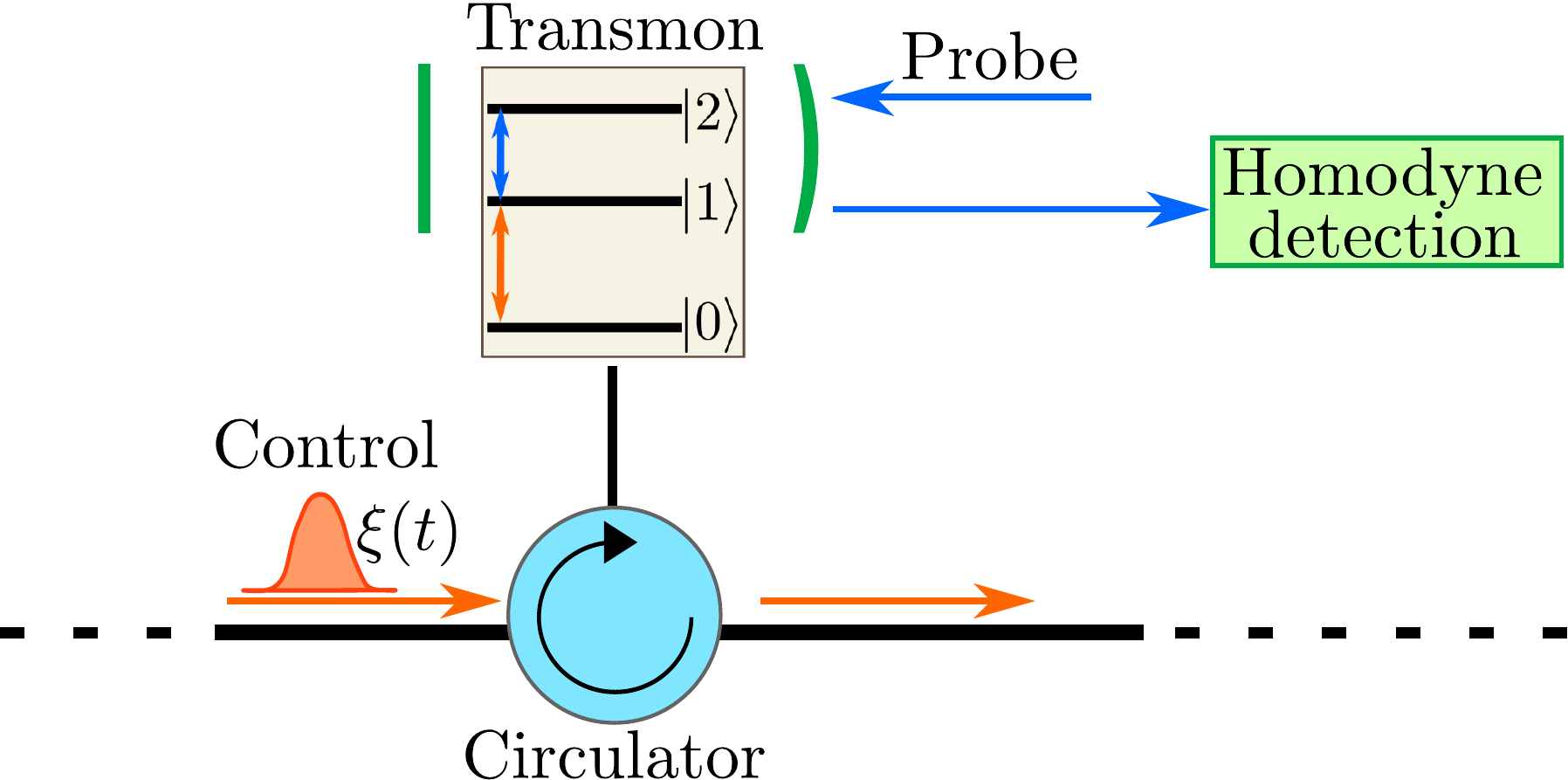}
 \caption{Schematic setup of a transmon with a cavity for the probe field. The setup can be extended by adding more units (transmon + cavity) using circulators to get a cascaded setup. The field reflected from each cavity is measured using a homodyne detector to infer the presence of the control photon.}
 \label{fig_transmons_w_cavity}
\end{figure}

As mentioned in the introduction, early experiments in QND detection of microwave photons were done in cavity QED. A cavity field can interact with the atom over several cycles, as the interaction time between the atom and the cavity field is much less compared to the cavity lifetime. However, if we are interested in detecting an incoming photon from arbitrary source (say from a different quantum node), we have to capture the photon in the cavity. To fully absorb the photon, the cavity needs to be of wide bandwidth and with strong coupling to the transmission line. High coupling however reduces the cavity lifetime, which degrades the performance of the photon detection schemes. However, as the probe field is only an auxiliary field for detecting the incoming control photon, one could imagine having a cavity for the probe only while the control photon is still an itinerant one. Such a setup (Fig. \ref{fig_transmons_w_cavity}) was proposed and analyzed in \cite{BixuanPRB2014}.

We will first consider a single unit comprising  of a transmon with a cavity for the probe field. In the rotating frame of the input fields, the Hamiltonian can be written including the coherent drive as
\be
H = \delta_1 \ket{1}\bra{1} + (\delta_1+\delta_2) \ket{2}\bra{2} - i E (b-b^\dagger) - i g (b \sigma_{21}-b^\dagger \sigma_{12})
\ee
where $b$ is the annihilation operator of the probe cavity, $E$ is the amplitude of the drive and $g$ is the coupling strength between the cavity and the $1-2$ transition. The stochastic master equation for this unit, using a cavity as a source of the control photon can be written as
\begin{align}
d\rho &= -i\comm{H}{\rho} + \left(\lind{L_{01}} + \lind{L_{12}} \right)\rho + \kappa_a(t) \lind{a} \rho +\kappa_b \lind{b} \rho - \sqrt{\kappa_a(t)}\; \coup{a}{L_{01}}\rho.
\end{align}
The field output from the probe cavity is measured using a homodyne detector and the corresponding homodyne current is
\be
j(t) dt = \sqrt{\kappa_b \eta} \expec{e^{-i\phi} b + e^{i\phi} b^\dagger} dt  + dW(t).
\ee
Similiar to previous section, the performance of this setup was analyzed numerically and the fidelity $F$ was shown to improve from 70\% to 84\%. This was achieved with the help of the cavity and by using an optimal linear filter $f(t)$ that takes the form of the expected homodyne current when the control field has a photon. The results also assume that the decay from the state $\ket{2}$ of the transmon into the transmission line can be suppressed to $\Gamma_{12}=0.1 \Gamma_{01}$. Lifting this restriction and using the usual transmon limit with $\Gamma_{12}=2 \Gamma_{01}$, the fidelity was found to be 81\%. Further improvements were shown to be possible by cascading the units similar to the previous section.








\section{Summary}
Microwave quantum optics using superconducting circuits, dubbed circuit QED, is an alternate approach for studying light-matter interaction that has been developing rapidly over the last few years. By confining the electromagnetic field to 1-dimension, strong coupling regime can be reached in these setups that has already led to plethora of  interesting results. A single photon detector in microwave regime will fill a major gap that would make these setups even more attractive for studying quantum optics and for developing quantum information technologies. The proposals we discussed in this article are aimed towards this goal. Especially, a QND single photon detector would lead to several interesting applications as discussed earlier. The setups (or similar ones) discussed for QND detection are close to experimental realization and will become even more attractive with on-chip circulators and quantum limited amplifiers. We believe that new and exciting physics in microwave quantum optics will be studied using single photon detectors in the near future.
\section*{Acknowledgements}


\begin{thebibliography}{00}
\bibitem{Einstein1905}
A. Einstein, Concerning an Heuristic Point of View Toward the Emission and Transformation of Light, Ann. Phys., 17, 132 (1905).

\bibitem{LoNatPhot2014}
H.-K. Lo, M. Curty, and K. Tamaki, Secure quantum key distribution, Nat Phot., 8, 595 (2014).

\bibitem{GiustinaNature2013}
M. Giustina et al., Bell violation using entangled photons without the fair-sampling assumption, Nature, 497, 227 (2013).

\bibitem{KnillNature2001}
E. Knill, R. Laflamme, and G. J. Milburn, A scheme for efficient quantum computation with linear optics, Nature, 409, 46 (2001).

\bibitem{BullerMST2010}
G. S. Buller and R. J. Collins, Single-photon generation and detection, Meas. Sci. Technol., 21, 12002 (2010).

\bibitem{EisamanRSI2011}
M. D. Eisaman et al., Invited Review Article: Single-photon sources and detectors, Rev. Sci. Instrum., 82, (2011).

\bibitem{MariantoniArxiv2005}
M. Mariantoni et al., On-Chip Microwave Fock States and Quantum Homodyne Measurements, Preprint, arXiv: cond-mat/0509737 (2005).

\bibitem{SilvaPRA2010}
M. P. da Silva et al., Schemes for the observation of photon correlation functions in circuit QED with linear detectors, Phys. Rev. A, 82, 43804 (2010).

\bibitem{BraginskySPU1975}
V. B. Braginskii and Y. I. Vorontsov, Quantum-mechanical limitations in macroscopic experiments and modern experimental technique, Soviet Physics Uspekhi, 17, 644 (1975).

\bibitem{BraginskyJETP1977}
V. B. Braginskii, Y. I. Vorontsov, and F. Y. Khalili, Quantum singularities of a pondero- motive meter of electromagnetic energy, J. Exp. Theor. Phys., 46, 705-706 (1977).

\bibitem{ThornePRL1978}
K. S. Thorne et al., Quantum nondemolition measurements of harmonic oscillators, Phys. Rev. Lett., 40, 667-671 (1978).

\bibitem{UnruhPRB1979}
W. G. Unruh, Quantum nondemolition and gravity-wave detection, Phys. Rev. B., 19, 2888-2896 (1979).

\bibitem{BraginskyScience1980}
V. B. Braginskii, Y. I. Vorontsov, and K. S. Thorne, Quantum nondemolition measurements, Science, 209, 547-557 (1980).

\bibitem{GrangierNature1998}
P. Grangier, J. A. Levenson, and J.-P. Poizat, Quantum non-demolition measurements in optics,  Nature,   396, 537 (1998).

\bibitem{SteanePRL96}
A. M. Steane, Error-correcting codes in quantum theory, Phys. Rev. Lett.,  77, 793 (1996).

\bibitem{RuskovPRB2003}
R. Ruskov, and A. N. Korotkov, Entanglement of solid-state qubits by measurement, Phys. Rev. B.,  67, 241305 (1993).

\bibitem{BishopNJP2009}
L.S. Bishop et. al, Proposal for generating and detecting multi-qubit GHZ states in circuit QED, New J. Phys.,  11, 073040 (2009).

\bibitem{RaussendorfPRL2001}
R. Raussendorf, and H. J. Briegel, One-way quantum computer, Phys. Rev. Lett.,  86, 5188 (2001).

\bibitem{SchusterNature2007}
D. I. Schuster et al., Resolving photon number states in a superconducting circuit,  Nature,   445,  515 (2007).

\bibitem{GuerlinNature2007}
C. Guerlin et al., Progressive field-state collapse and quantum non-demolition photon counting, Nature,   448, 889 (2007).

\bibitem{WangPRL2008}
H. Wang et al., Measurement of the decay of a Fock states in a superconducting circuit, Phys. Rev. Lett.,  101, 240401 (2008).

\bibitem{JohnsonNature2010}
B. R. Johnson et al., Quantum non-demolition detection of single microwave photons in a circuit,  Nat. Phys., 6, 663 (2010).

\bibitem{YinPRL2013}
Y. Yin et al., Catch and Release of Microwave Photon States, Phys. Rev. Lett., 110, 107001 (2013).

\bibitem{WennerPRL2014}
J. Wenner et al., Catching Time-Reversed Microwave Coherent State Photons with 99.4\% Absorption Efficiency, Phys. Rev. Lett., 112, 210501 (2014).

\bibitem{FlurinPRL2015}
E. Flurin et al., Superconducting Quantum Node for Entanglement and Storage of Microwave Radiation, Phys. Rev. Lett., 114, 90503 (2015).

\bibitem{RomeroPRL2009}
G. Romero, J. J. Garc\'ia-Ripoll, and E. Solano, Microwave Photon Detector in Circuit QED, Phys. Rev. Lett. 102, 173602 (2009).

\bibitem{RomeroPhysScr2009}
G. Romero, J. J. Garc\'ia-Ripoll, and E. Solano, Photodetection of propagating quantum microwaves in circuit QED, Phys. Scr. 2009, 14004 (2009).

\bibitem{PeropadrePRA2011}
B. Peropadre et al., Approaching perfect microwave photodetection in circuit QED,  Phys. Rev. A.,   84, 063834 (2011)

\bibitem{ChenPRL2011}
Y. F. Chen et al., Microwave photon counter based on josephson junctions,  Phys. Rev. Lett.,   107,  217401 (2011).

\bibitem{PoudelPRB2012}
A. Poudel, R. McDermott, and M. G. Vavilov, Quantum efficiency of a microwave photon detector based on a current-biased Josephson junction, Phys. Rev. B, 86, 174506 (2012).

\bibitem{GoviaPRA2012}
L. C. G. Govia et al.,  Theory of Josephson photomultipliers: Optimal working conditions and back action, Phys. Rev. A, 86, 32311 (2012).

\bibitem{GoviaPRA2014}
L. C. G. Govia et al., High-fidelity qubit measurement with a microwave-photon counter, Phys. Rev. A, 90, 62307 (2014).

\bibitem{KochPRA2007}
J. Koch et al., Charge-insensitive qubit design derived from the Cooper pair box,  Phys. Rev. A.,   76, 042319 (2007).

\bibitem{IoChunPRL2013}
I.-C. Hoi et al., Giant cross Kerr effect for propagating microwaves induced by an artificial atom, Phys. Rev. Lett.,   111, 053601 (2013).

\bibitem{BixuanPRL2013}
B. Fan et al., Breakdown of the cross-Kerr scheme for photon counting, Phys. Rev. Lett.,   110, 053601 (2013).

\bibitem{SankarPRL2014}
S. R. Sathyamoorthy et al., Quantum Nondemolition Detection of a Propagating Microwave Photon, Phys. Rev. Lett., 112, 093601 (2014).

\bibitem{BixuanPRB2014}
B. Fan et al., Nonabsorbing high-efficiency counter for itinerant microwave photons. Phys. Rev. B.,   90, 035132 (2014)

\bibitem{Gough2012}
J. E. Gough, M. R. James, H. I. Nurdin, and J. Combes. Quantum filtering for systems driven by fields in single-photon states or superposition of coherent states, Phys. Rev. A., 86, 043819 (2012).

\bibitem{Lax66}
M. Lax, Quantum Noise. IV. Quantum Theory of Noise Sources, Phys. Rev. 145 110 (1966)

\bibitem{CarmichaelQuantumOptics}
H. J. Carmichael, An Open Systems Approach to Quantum Optics, Springer-Verlag, (1993)

\bibitem{GoughCommMathPhys2009}
J. Gough, and M. R. James. Quantum Feedback Networks: Hamiltonian Formulation, Commun. Math. Phys.  287, 1109 (2009)

\bibitem{GoughIEEE2009}
J. Gough, and M. R. James. The Series Product and Its Application to Quantum Feedforward and Feedback Networks, IEEE Trans. Autom. Control,  54, 2530(2009).

\bibitem{KochPRA2010}
J. Koch et al., Time-reversal-symmetry breaking in circuit-QED based photon lattices. Phys. Rev. A.,  82, 043811 (2012)

\bibitem{KerckhoffArxiv2015}
J. Kerckhoff et al., On-chip superconducting microwave circulator from synthetic rotation. Preprint. arXiv:1502.06041 (2015)

\bibitem{SliwaArxiv2015}
K. M. Sliwa et al., An integrated Josephson circulator and directional amplifier: the triple-pumped JPC. Preprint. arXiv:1503.00209  (2015)

\bibitem{StacePRL2004}
T. M. Stace, C. H. W. Barnes, and G. J. Milburn, Mesoscopic One-Way Channels for Quantum State Transfer via the Quantum Hall Effect, Phys. Rev. Lett. 93, 126804 (2004).

\end{thebibliography}
\end{document}